# Newton's 2nd Law and the Physics of Dance

Richard P. Barber, Jr.
Department of Physics, Santa Clara University, Santa Clara, CA 95053

David J. Popalisky
Department of Theatre and Dance, Santa Clara University, Santa Clara, CA 95053

Rose Hacking and Kristina Chiapella
Santa Clara University, Santa Clara, CA 95053

## Abstract

In teaching the physical sciences, a significant challenge lies in the student's tendency to consider the scientific world and the "real" world as separate. For example, Newton's $1^{st}$ Law of Motion states that an *object* in motion remains in motion in a straight line unless acted on by an external force. However, our experience tells us that most objects keep moving only as long as someone or something pushes on them. One key to understanding physics is the ability to abstract the "law" from a reality which also includes friction and other effects. In this article we describe a college course for non-science majors, *The Physics of Dance*. The central theme of this course is the personalization of the physics of motion by making each student the *object*. With this approach we give students not only scientific tools to measure and understand but personal involvement to experience forces and motion. This combination provides a bridge that connects the science to reality.

Santa Clara University's Core Curriculum laboratory requirement in the natural sciences expects that "through an introduction to the scientific method based as much as feasible on actual laboratory exercises, students should demonstrate the ability to raise questions about science and technology in their own lives, as well as in the society in which they live." In this article we will describe *The Physics of Dance*,[1] a course that fulfills this requirement. In addition to outlining the class, we will focus on experiments from the course that directly explore Newton's Second Law of motion. Finally we will present an example of a student project which is evidence for the effectiveness of this approach.

In this course we use discussion, lecture and laboratory exercises to engage students in the learning process. Both an experimental record and an experiential reflection are included in a laboratory notebook. The point of this "dual" record is to help students connect personal experience and sensation (the dance) with objective scientific observation (the physics). The course developed from an interdisciplinary conversation that illuminated the similarities between the creative and analytical processes that are common to both physics and dance. We are convinced that imbedding this connection in the class is critical to its success.

In order to describe the course organization and approach, we will present an outline of a typical week. Generally the main course topics are covered in one-week sections. For our example we have chosen vertical jumps, and each component of the class will be presented within this context. The course is taught as a Tuesday-Thursday lecture/discussion in a physics classroom with a laboratory on Wednesday in a dance studio. We introduce the concept(s) for the week first through the reading assignment due by class on Tuesday. The text, *Physics and the Art of* Dance,[2] is a semi-quantitative presentation of the physical laws of motion as expressed through the language of classical ballet. A three-question quiz is given at the beginning of class to begin our discussion. The quiz discussion leads into the central dance concepts for the week. At that point we begin to develop simple physical derivations to show how we model the fundamental dance ideas. In this approach we start by introducing the dance and then slowly adding the physics. For example when we discuss Newton's Second Law, we consider what forces act *on* a solo dancer. These pushes and pulls are simply gravity, the vertically upward force of the floor and horizontal friction forces from the floor.

For our vertical jump example our derivation simplifies because we have eliminated any horizontal forces. During the Tuesday lecture we introduce the elementary physics by determining that the net force leads to the following expression:

$$F_{NET} = F_{floor} - mg = ma.$$

In a week previous to this discussion, the class has already covered motion during constant acceleration. It is straightforward to show that while the dancer's feet are off the floor, the value of the constant acceleration is $-g$. Since the floor does not act in this case, we easily find this result from

$$F_{NET} = -mg = ma \text{ so that}$$

$$a = -g.$$

We do not attempt a detailed description of the forces and acceleration during the takeoff and landing of the jump at this point. We will revisit that discussion once we have acquired laboratory data that will allow us to probe the more subtle details of the forces and motion.

For the Wednesday lab, we start with a warm-up before introducing choreographed movements specific to the week's topic. For our vertical jump example there is particular focus on warming up legs and ankles to both increase the efficiency of the jumps and to avoid injuries. Once students have had time to perform the particular movement, they are given some time to reflect on the aesthetic experiences by writing in their laboratory journals. For the last part of the laboratory, the class breaks into pairs to conduct their own experiments. This work includes specific assigned measurements as well as experiments on variations of the dance movements that they create on their own. Their results are also recorded in the laboratory notebooks. Over the course of the laboratory session we have again started with the dance concepts and added the experiments towards the end.

The class uses two measurement tools – digital video cameras from which we can extract position vs. time data, and force plates from Vernier Software & Technology that are similar to bathroom scales but read force vs. time at 50 readings/second. With the digital video, we use a calibrated background or a meter stick in the field-of-view to provide a length scale. Data are extracted by hand using a ruler on a computer screen to measure frame-by-frame position. The actual distances are then scaled using the calibration from the background or meter stick. The force plates are recorded using a handheld interface/data logger and software provided by Vernier. Except for zeroing the force plates before an experimental run, these devices require no further setup. For our vertical jump example, we will use videotape and force plate data recorded simultaneously.

The class on Thursday focuses on the laboratory results from the previous day. We discuss the lab activities and how the data produced connect with the physics that we introduced on Tuesday. The students have recorded both experimental data and their own kinesthetic experience to facilitate this discussion. For our example of vertical jumps, we explore both the motion and the forces that produce it. We have a computer in class to show the students how to do the analysis in LoggerPro, the Vernier Software acquisition, plotting, and analysis package. We do a frame-by-frame analysis of the video clip of the movement and demonstrate converting the image into numerical position, velocity and acceleration data. Although the software supports doing this video position analysis automatically, we have the students find the position in each frame by hand. Given that our student population consists primarily of non-science majors, we feel that it is essential to show this analysis in real-time without too many layers of automation in order for them to develop basic skills and understanding. The force plate data are often recorded using two force plates. With one plate under each foot, this arrangement provides a safer and more stable target on which students can jump. This approach also requires a very simple analysis: adding two force plate data columns. Later in this article, we will discuss how one student team used this approach to uncover interesting results. Since Thursday is primarily focused on data analysis, we spend significant time making graphs, rescaling the data, doing simple slope calculations etc. As the students learn and build these skills, we leave out steps in subsequent weeks.

Figure 1 presents our primary data for a single vertical jump. These results were recorded using simultaneous videotaping and force plate readings. Shown are the vertically upward force vs. time and the height in meters vs. time. These latter data have already been rescaled to reflect the correct length calibration for the room. The students are shown how to find the position in *cm* on the computer screen and then rescale those data into *m* in the room.

Figure 2 focuses on the motion data from figure 1. We have three views, the height vs. time, the numerically calculated vertical velocity vs. time, and the calculated vertical acceleration vs. time. The calculated results are derived from the position data and the $1/30^{th}$ second time intervals given by the 30 frames per second rate of the digital video cameras. The roughly linear region in the center of the velocity graph (the free-fall region) has been fit for a slope. The value of -10.1 m/s$^2$ is the estimate of *g* from these data which compares well with the known value of -9.7996 m/s$^2$ in Santa Clara, CA.[3] It is important to point out that simple techniques are favored for calculating the slopes from the data instead of more sophisticated ones. In class these calculations and plots are made in real time, so the impact of discovery and understanding is enhanced by the speed of simple calculations.

The bottom frame of figure 2 displays a rather rough but notable estimate of the acceleration during the jump. Students in the class easily recognize that this curve resembles force plate data (figure 1a). In figure 3 we make a direct comparison by showing the product of the calculated student acceleration and mass compared to the net force. This latter quantity is derived by noting that the force plates simply measure the upward force on the student. If the jumper stands still, her weight is measured. This quantity can be subtracted from the upward force vs. time to yield net force vs. time as we display. Although the "noisiness" from the slope calculation is large, the direct comparison is quite compelling and serves as a direct visual connection to Newton's Second Law:

$$F_{NET} = ma.$$

Each week we attempt to build on the experiments and analysis of the previous week, steadily increasing skills and sophistication. In lieu of a final exam the course concludes with a final project: a simple choreography, components of which are recorded and analyzed using the set of tools developed over the quarter. This is a culminating opportunity where students are acquainted with the parallel activities of analysis and repetition that characterize both the creative and the experimental process. In the final presentations the student groups first perform their choreography; then they give a brief talk on their analysis and conclusions; and finally they perform the dance again. Our intent is to give the students enough experience with simple choreographic craft, devising experiments, recording data, analyzing results, and making conclusions so that they can present creative, innovative, and scientifically sound presentations for this final work.

As an example of work from such a project, we present some student results. During the quarter we discuss impulse and momentum, because impulse is easy to calculate using the force plate data. Performing some preliminary analysis of vertical jumps, Rose and Kristina noticed asymmetries in left foot-right foot impulses as they were performing jumps. As we pointed out in our earlier discussion, this analysis was made possible because two force plates were used to record the data. Instead of simply adding the two data sets immediately, Rose and Kristina noted differences that became a focus of their final project. In figure 4 we see two typical jumps by Kristina. The top frame is a left foot force measurement, the middle is the right foot force, and the bottom is the sum. These curves have been subtracted to yield the "net" force on each foot, and the positive impulse of both the takeoffs and landings have been computed and labeled. Note that the impulse beginning the jump is dominated by the left foot where the right foot dominates the landing. This observation suggests that perhaps Kristina is tilting during the jump. The total impulses for the takeoff and landing are comparable as we would expect given the same change in momentum for both cases. This last result gives us some confidence that the measurements are reliable. For jumps by Rose, we observe a different result in figure 5. In her case the right foot provides the most impulse on both takeoff and landing, suggesting that she is right-footed or one leg is longer than the other. Another explanation would be an imbalance in her muscular stretch or strength that would result in a right-leg dominance. Again the total impulses for takeoff and landing on each jump compare reasonably well, so we have confidence in the measurements.

The value in the student projects is largely pedagogical. In the case of the vertical jumps and asymmetric impulses, there might also be some interest for professional dancers in fine-tuning their technique for performance and efficiency. Such asymmetries might suggest the need for improvement in the aesthetics of jumps, as well as elucidating potential injury-prevention techniques. This example might also rise to the level of a student research project with more sophisticated analysis, detailed modeling, and more extensive data sets with more human subjects. Such projects have provided successful undergraduate research experiences.[4] Our point here is to provide evidence of student learning and understanding through this example of a student-conceived and analyzed experiment.

In conclusion, *The Physics of Dance* involves students directly and personally in the study of physics. We have presented a brief overview of the course and given an example of one topic covered during a typical week. An example student project provides evidence of the effectiveness of our approach. We are convinced that the kinesthetic experience enhances student understanding and makes getting a glimpse at the scientific process more fruitful. By blending science with a performing art, our intent is to draw non-science students into the scientific approach while honoring both the creative and analytical aspects of both disciplines.

Course development was funded by two Santa Clara University Technology Innovation Grants and a Thomas Terry Grant. We also acknowledge helpful discussions with Phil

Kesten, Ken Laws, Norm Hirschy, Arleen Sugano, and the Yockeys. Thanks to Hermione Sharp and Tim Ross for vertical jump data.

# References

[1] We are aware of two other courses that make a similar connection: *Physics of Dance* developed by Pan Papacosta at Columbia College, Chicago and *Physics of Movement* developed by Andrea Lommen at Franklin and Marshall College.

[2] K. Laws, *Physics and the Art of Dance* (Oxford University Press, New York, 2002).

[3] Value given on plaque in former seismograph location in Ricard Memorial Observatory, Santa Clara University, Santa Clara, CA.

[4] M. Cluss, K. Laws, N. Martin, T. S. Nowicki, and A. Mira, "The indirect measurement of biomechanical forces in the moving human body," *Am. J. Phys.* **74**, 102 (2006).

**Figure Captions**

Fig 1. a) Force vs. time from a force plate reading of a vertical jump. b) Vertical height vs. time from videotape date of the same jump.

Fig 2. Results from the vertical jump shown in fig. 1 plotting a) vertical height vs. time, b) vertical velocity vs. time, and c) vertical acceleration vs. time. The linear fit shown in the center of b) gives an estimate for *g* of -10.1 m/s$^2$.

Fig 3. An overlay of the student's mass times the calculated vertical acceleration from fig. 2 with the force vs. time plot from fig. 1.

Fig 4. Force vs. time and calculated impulses for two vertical jumps by Kristina showing a) left foot, b) right foot, and c) total force.

Fig 5. Force vs. time and calculated impulses for two vertical jumps by Rose showing a) left foot, b) right foot, and c) total force.

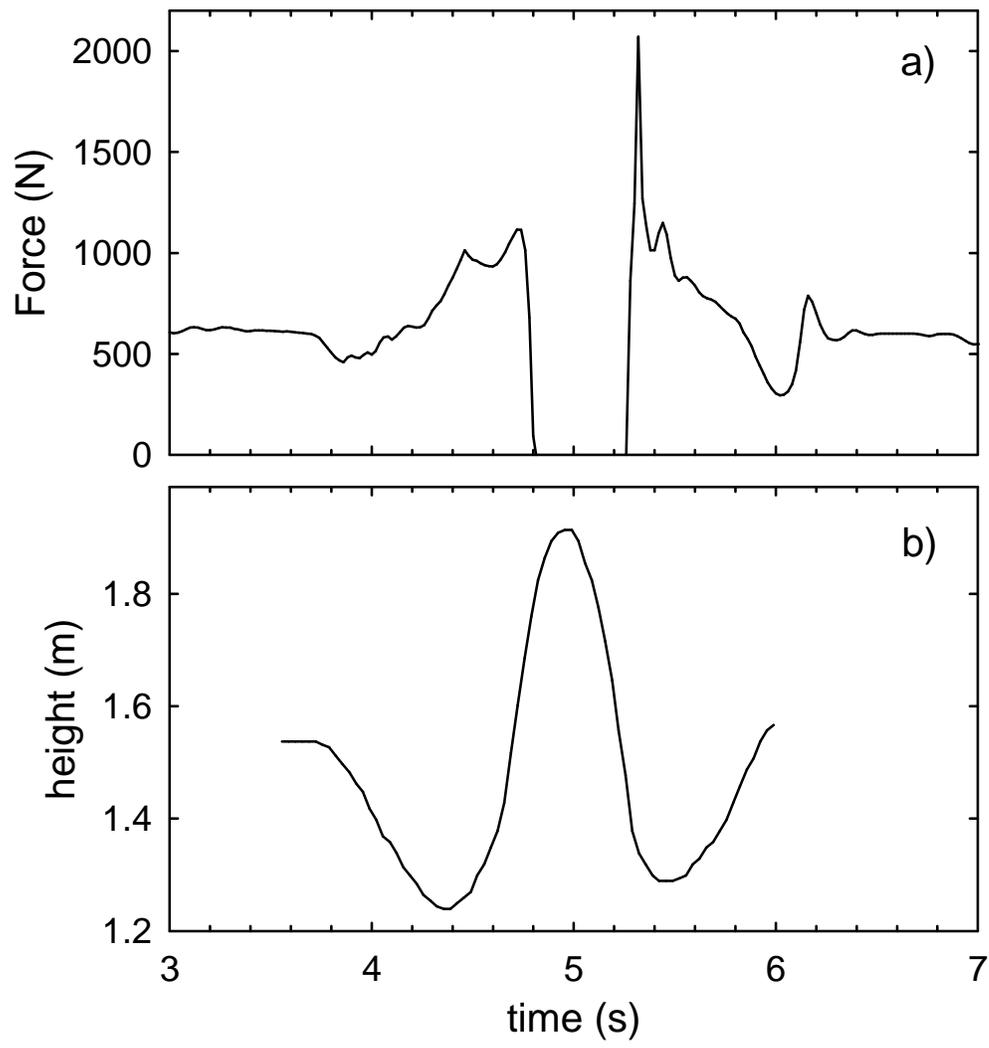

Fig 1

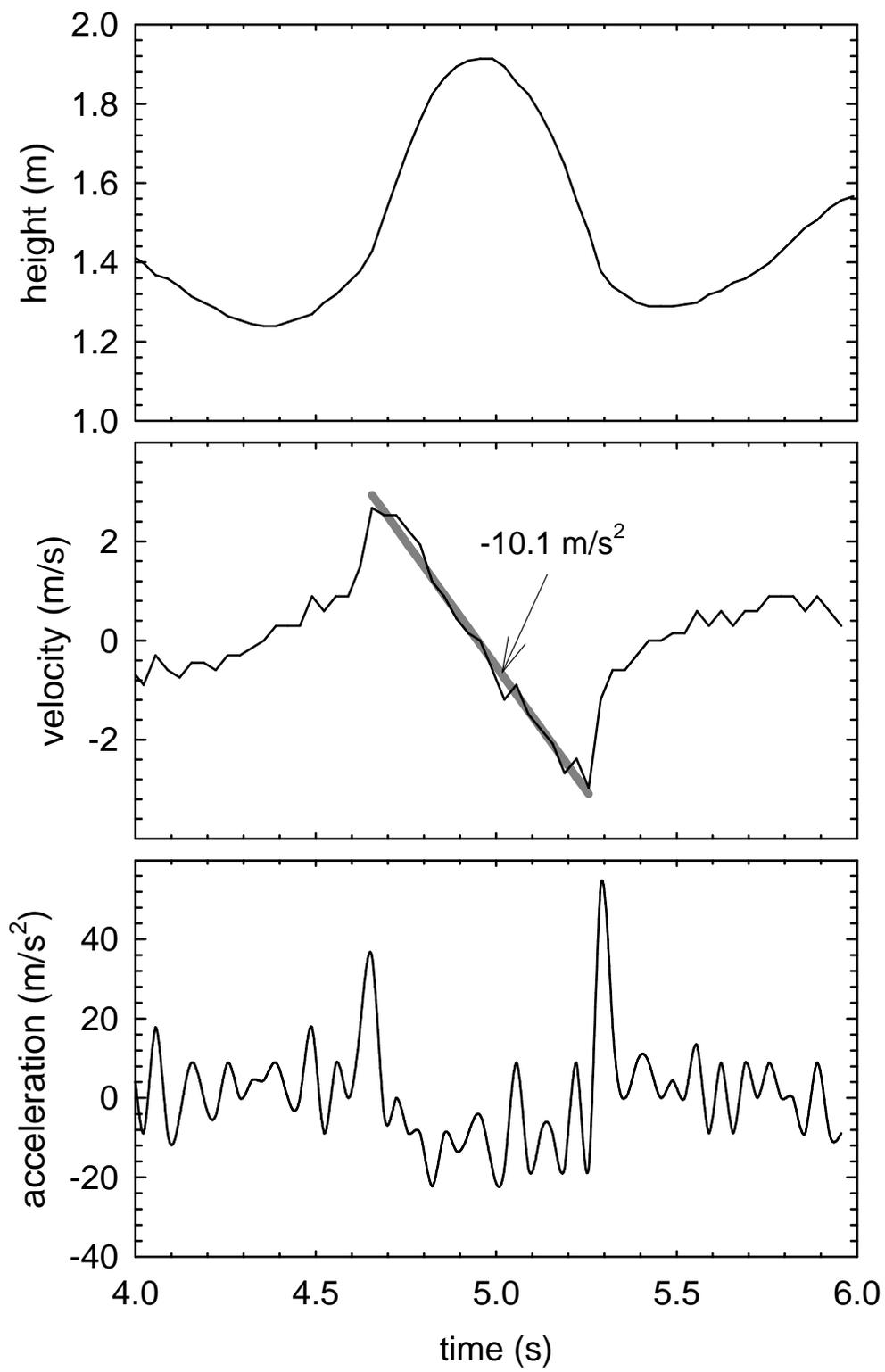

Fig 2

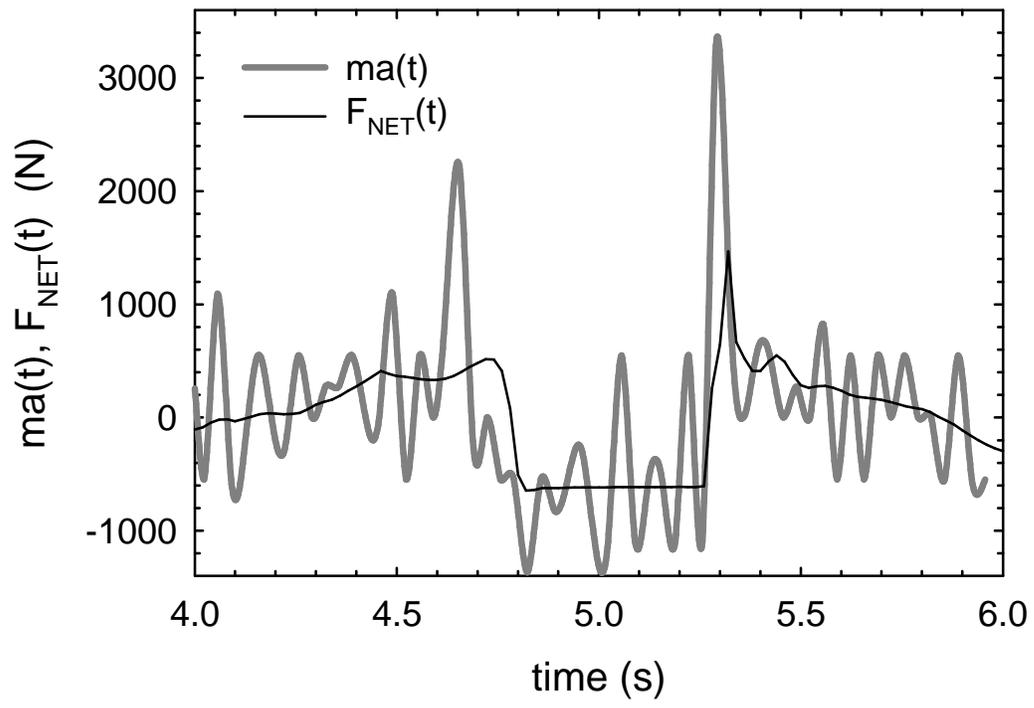

Fig 3

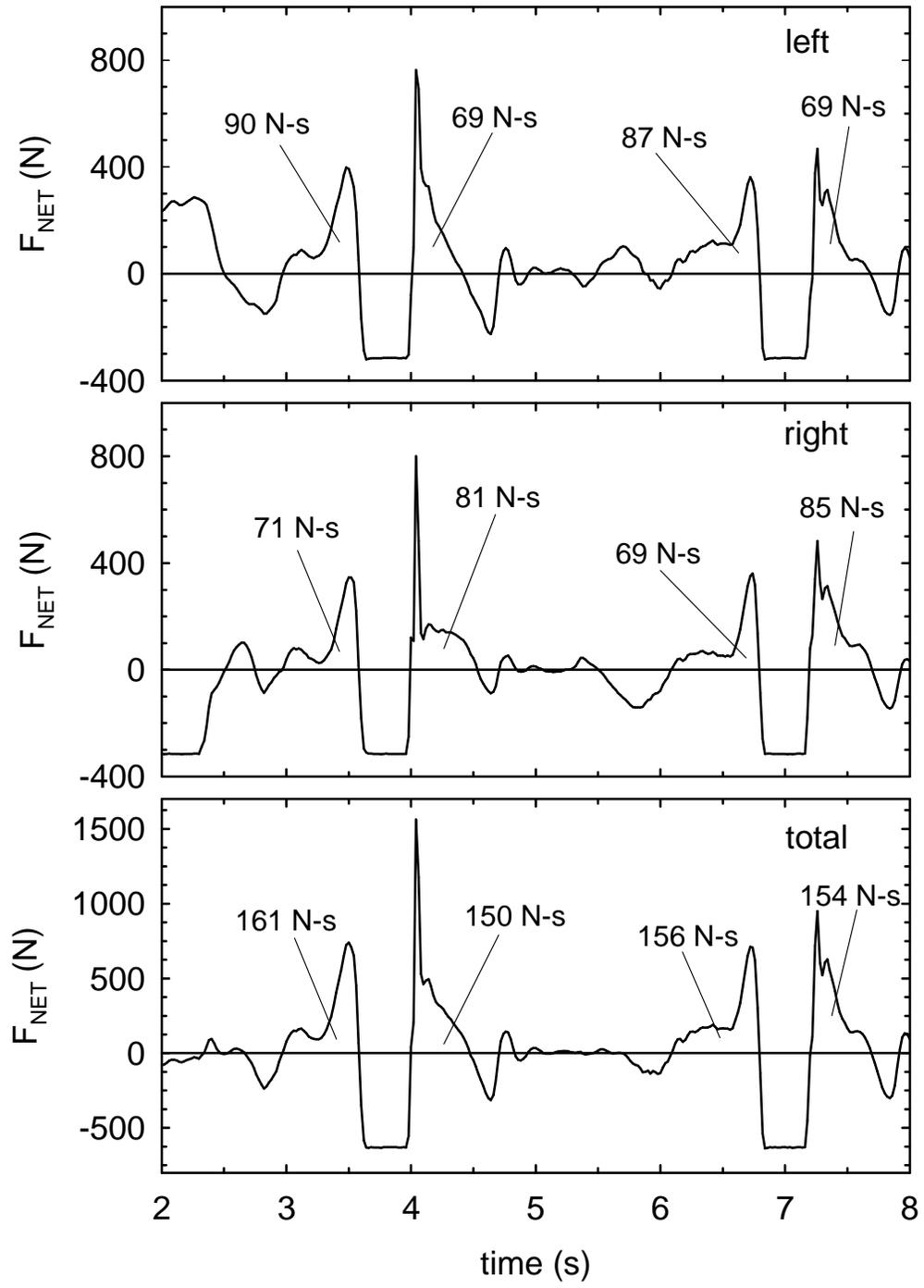

Fig 4

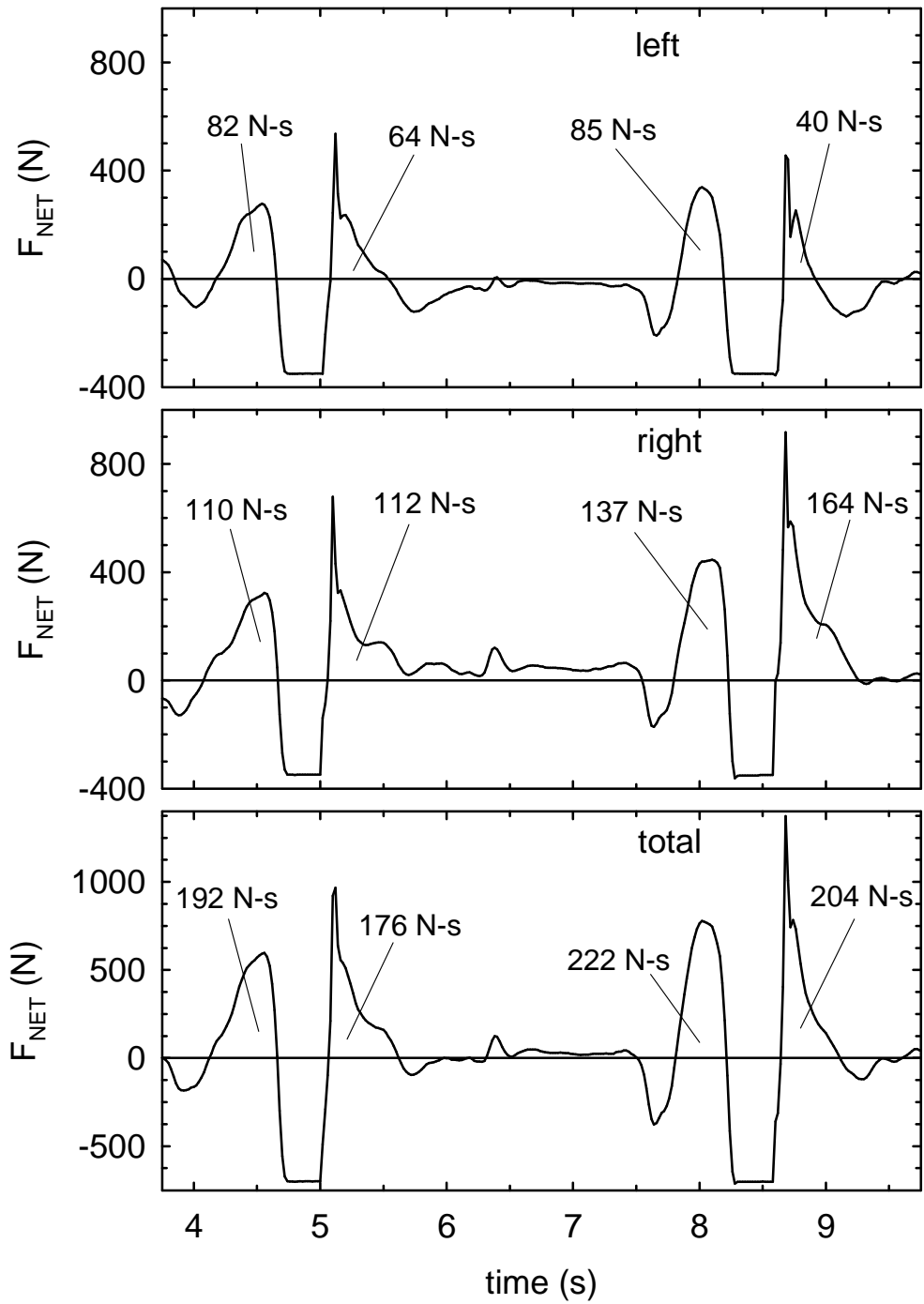

Fig 5